\pdfoutput=1

\documentclass[aps,prb,reprint,showpacs,superscriptaddress]{revtex4-1}

\usepackage{graphicx}
\usepackage{graphics}
\usepackage{amsmath}
\usepackage{amssymb}
\usepackage{amsfonts}
\usepackage{dcolumn}
\usepackage{dsfont}
\usepackage{latexsym}
\usepackage{rotating}
\usepackage{color}
\usepackage{latexsym}
\usepackage{bbm}
\usepackage{subfigure}
\usepackage{float}
\usepackage{epsfig}
\usepackage{epsf}
\usepackage{psfrag}
\usepackage{bm}
\usepackage{amsthm}
\usepackage{eucal}
\usepackage{mathrsfs}
\usepackage{url}
\usepackage{braket}

\usepackage{color} 

\usepackage{hyperref}
\hypersetup{
colorlinks=true,final=true,
        linkcolor=blue,
        citecolor=blue,
        filecolor=blue,
        urlcolor=blue,
}
\begin{document}
\title{Electronic structure and magnetism in infinite-layer nickelates RNiO$_2$ (R= La-Lu)}

\author{Jesse Kapeghian}
\affiliation{Department of Physics, Arizona State University, Tempe, AZ - 85287, USA}

\author{Antia S. Botana}
\email{antia.botana@asu.edu}
\affiliation{Department of Physics, Arizona State University, Tempe, AZ - 85287, USA}
\date{\today}
\begin{abstract}

{Using first-principles calculations, we analyze the evolution of the electronic structure and magnetic properties of infinite-layer nickelates RNiO$_2$ (R= rare-earth) as R changes across the lanthanide series from La to Lu. By correlating these changes with in-plane and out-of-plane lattice parameter reductions, we conclude that the in-plane Ni-O distance is the relevant control parameter in infinite-layer nickelates. An antiferromagnetic ground state is obtained for all RNiO$_2$ (R=La-Lu). This antiferromagnetic state remains metallic across the lanthanide series and is defined by a multiorbital picture with low-energy relevance of a flat Ni-d$_{z^2}$ band pinned at the Fermi level, in contrast to cuprates. Other non-cuprate-like properties such as the involvement of R-$d$ bands at the Fermi level and a large charge transfer energy are robust for all  RNiO$_2$ materials. 
}

\end{abstract}
\maketitle
\section{Introduction}

The collection of characteristics sought for when looking for cuprate analogs include a layered structure, proximity to a d$^9$ (S=1/2) configuration analog to Cu$^{2+}$, d$_{x^2-y^2}$ states as the active orbitals, antiferromagnetic correlations, and strong p-d hybridization \cite{norman-RPP}. Perovskite RNiO$_3$-based heterostructures (R = rare-earth) have been intensively studied over the last decade motivated by predictions of superconductivity based on their analogies to cuprates \cite{held, khaliullin, review_113, review_113_2, review_113_3}.  A plethora of new phenomena have been discovered in these systems, due to the interplay of epitaxial strain, quantum confinement, and interfacial effects. However, the promise of RNiO$_3$-based heterostructures for superconductivity is yet to be realized \cite{review_113, review_113_3}. 

In this regard, one alternative to heterostructures based on perovskite phases are low-valence layered nickelates R$_{n+1}$Ni$_n$O$_{2n+2}$ (n= 2, 3... $\infty$), closer to cuprates in terms of their structure (with infinite NiO$_2$ planes) as well as in terms of electron count (close to d$^9$)\cite{poltavets1, poltavets2, nat_phys, physrevmat}.  The realization of this promise came with the recent report of the first superconducting nickelate: hole-doped  NdNiO$_2$ \cite{new, sc_1, sc_2}. RNiO$_2$ (112) materials are the infinite-layer members of the series and realize the hard to stabilize Ni$^{1+}$ oxidation state, isoelectronic with Cu$^{2+}$ \cite{anisimov}. These materials are derived via oxygen reduction from the corresponding perovskite phase RNiO$_3$ (113) \cite{hayward_nd, hayward, crespin, ikeda, ikeda2}. 
To date, only the Nd and La variants of 112 materials have been realized in bulk form \cite{hayward_nd, hayward, crespin, ikeda, ikeda2}. However, parent perovskite 113 phases exist for R=Lu-La, and their phase diagram is well established \cite{review_113_2}.
Except for R = La, all rare-earth perovskite 113 nickelates exhibit a
metal–insulator phase transition (MIT),
accompanied by a symmetry lowering from orthorhombic to monoclinic \cite{review_113, review_113_2}. At a temperature lower than T$_{MIT}$, they undergo an antiferromagnetic (AFM) phase transition. The evolution
of these transitions can be correlated with structural changes upon a change in rare-earth. In particular, the Goldschmidt tolerance factor (that serves as a measurement of the tendency of the structure to distort) is often advocated in this regard- a decrease in tolerance
factor with R tends to reduce the Ni–O–Ni angle with a
subsequent reduction of the overlap between the Ni-d and O-p orbitals \cite{review_113_2, review_113_3}.

\begin{table}
\caption{In-plane ($a$) and out-of-plane ($c$) ab-initio optimized lattice parameters for RNiO$_2$ (R= rare-earth)  within GGA.}
\begin{ruledtabular}
\begin{tabular}{lcc}
\multicolumn{1}{l}{Rare-earth} &
\multicolumn{1}{l}{a (\AA)} &
\multicolumn{1}{c}{c (\AA)} \\

   \hline
  
          La  & 3.960 & 3.370 \\ 
           Pr   & 3.940 & 3.341 \\ 
            Nd          & 3.926 & 3.302 \\ 
             Pm        	&  3.912 & 3.263 \\ 
                       Sm       	&  3.902 & 3.237 \\ 
        Eu  & 3.890 & 3.206 \\ 
        Gd  & 3.879 & 3.177 \\ 
        Tb  & 3.869 & 3.155 \\ 
        Dy & 3.861 & 3.133 \\ 
        Ho  & 3.856 & 3.110 \\ 
        Er  & 3.849 & 3.090 \\ 
        Tm  & 3.841 & 3.070 \\ 
        Yb  & 3.834 & 3.053 \\ 
        Lu  & 3.828 & 3.034 \\

\end{tabular}
\end{ruledtabular}
\label{table1}
\end{table}

\begin{figure*}
\includegraphics[width=2\columnwidth]{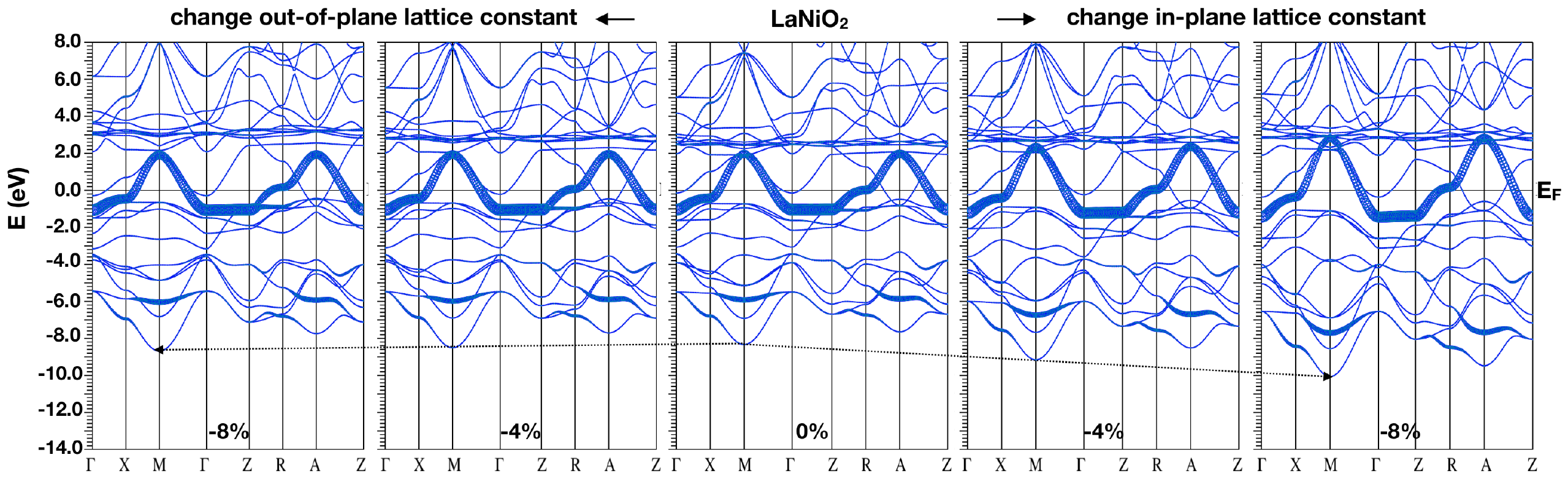}
\caption{GGA non-magnetic LaNiO$_2$ band structure with band character plot (Ni-d$_{x^2-y^2}$ highlighted) at the experimental lattice constants (0\%, central panel), upon reduction of in-plane (right panels) and out-of-plane (left panels) lattice parameters. The arrows are a guide to show the drastic changes in the bandstructure upon an in-plane lattice constant reduction in contrast to the negligible changes upon an out-of-plane lattice constant reduction.  Appendix Fig. \ref{figa2} shows a zoomed-in version of the band structures around the Fermi level.}
\label{fig1}
\end{figure*}

Exploiting the prospect that other perovskite nickelate 113 phases could be reduced to their respective 112 counterpart, we analyze here the electronic structure and magnetism of RNiO$_{2}$ materials (R= La-Lu). We correlate changes in R with modifications of the in-plane and out-of-plane lattice parameters in LaNiO$_{2}$ (the extreme member of the lanthanide series) and analyze the effects of these changes in their electronic and  magnetic properties. We conclude that the Ni-O in-plane distance is the control parameter in RNiO$_{2}$ nickelates as modifications of the out-of-plane lattice parameter do not give rise to appreciable changes in the electronic structure. We find that all infinite-layer nickelates have an antiferromagnetic ground state from first principles, but this state is fundamentally different from that of cuprates as it is metallic and has Ni-d$_{z^2}$ states as active orbitals (in addition to the naively expected d$_{x^2-y^2}$ ones).

\section{computational details}

We performed two separate sets of density functional theory (DFT)-based calculations for RNiO$_2$ nickelates. 1) \textit{Calculations with a different rare-earth ion}. We performed calculations for 112 compounds across the lanthanide series (R= La-Lu) using PAW pseudopotentials\cite{KressePAW} as implemented in the VASP code \cite{VASP1,VASP2} placing the R-4f electrons in the core. We note that the role of 4f states in NdNiO$_2$ has been analyzed in the literature in connection to its normal and superconducting state properties\cite{pickett2}. It would be worth analyzing what aspects  of the electronic structure and magnetic properties of  RNiO$_2$ materials may be affected from the open 4f shell for other R ions but we leave that for future work. 2) \textit{Calculations for LaNiO$_2$ at different in-plane and out-of-plane lattice constants}. These calculations were  performed using the all-electron, full potential code WIEN2k~\cite{wien2k, SCHWARZ2003259} based on the augmented plane wave plus local orbitals (APW + lo) basis set keeping the La-4f states present. As a starting point, the experimental LaNiO$_2$ lattice constants were used \cite{hayward} ($a$= 3.96 \AA, $c$= 3.37 \AA). Using reduced lattice parameters in LaNiO$_2$ allows us to mimic the effect of a smaller R ion (the size of R  decreases with increasing atomic number) and determine if the changes in  electronic structure and magnetic properties upon R variation are linked mostly to the explicit R change, or to in-plane ($a$) and/or out-of-plane ($c$) lattice constant changes. 

The Perdew-Burke-Ernzerhof version of the generalized gradient approximation (GGA)~\cite{pbe} was used for both WIEN2k and VASP structural relaxations, and for non-magnetic calculations. 
In order to properly account for correlation effects in the Ni-d electrons, an on-site
Coulomb repulsion U was included in spin-polarized calculations, using the GGA+U method within the fully localized limit (FLL) \cite{LiechPRB}. For both VASP and WIEN2k GGA+U calculations, we used U-values ranging from 1.4 to 6.4 eV.  A nonzero value of Hund's coupling J= 0.7 eV has been considered to account for
the anisotropy of the interaction \cite{pickett_u}. Different magnetic configurations were checked: a) a ferromagnetic (FM) order, b) a C-type AFM order for which a $\sqrt{2}$ $\times$ $\sqrt{2}$ cell  was used, and c) a G-type AFM order for which a $\sqrt{2}$ $\times$ $\sqrt{2}$ $\times$ 2 cell was constructed. Given that a C-type AFM order is more stable for LaNiO$_2$ at a reasonable U value for this metallic nickelate $\sim$ 4-5 eV\cite{Sakakibara_arxiv2019, arita}  (see Appendix Fig. \ref{figa1}), a C-type AFM order was adopted for the other R112 materials for a systematic comparison. We note that a C-type AFM order is also found to be the groundstate in Ref. \onlinecite{liu} using GGA+U with FLL as the double-counting correction. 

In VASP, the wave functions were expanded in the plane-wave basis with a kinetic energy cutoff of 500 eV. The reciprocal space integration was carried out with a 16$\times$16$\times$16  $\Gamma$-centered k-mesh  for non-magnetic calculations, and a 12$\times$12$\times$16  $\Gamma$-centered k-mesh for antiferromagnetic calculations.
For all WIEN2k calculations, an RK$_{max}$ of 7 was used, as well as a k-mesh of 20$\times$20$\times$23 for non magnetic calculations and 18$\times$18$\times$30 for antiferromagnetic calculations. The muffin-tin radii used in WIEN2k for LaNiO$_2$ are 2.50 \AA~ for La, 1.99 \AA~ for Ni and, 1.72 \AA~ for O.

\section{results}

\subsection{Structural properties of RNiO$_2$}

Experimentally reported RNiO$_2$ (R= La, Nd) nickelates have a tetragonal structure with P4/mmm space group with $a=b$ $\neq$ $c$ \cite{hayward_nd, hayward}. The R, Ni and O positions are (0.5, 0.5, 0.5), (0,0,0), and (0.5, 0, 0), respectively. Using this data, we construct the structures for the hypothetical RNiO$_2$ materials whose lattice parameters were subsequently optimized for different R ions within GGA in a $\sqrt{2}\times\sqrt{2}$ cell with C-type AFM order. The corresponding lattice constants are shown in Table \ref{table1} and are in agreement with those reported in Ref. \onlinecite{Been2020TheoryOR}. Cerium was skipped due to its stable 4+ oxidation state. As expected, the lattice parameters decrease as the size of the rare-earth atom decreases with increasing atomic number. The out-of-plane lattice constant ($c$) gets reduced by 10\% across the lanthanide series from La to Lu, the in-plane lattice constant ($a$) by $\sim$ 4\%. We note that the Ni-O in-plane distance corresponds to a/2. The R-O distances are in agreement with the sum of ionic radii for the corresponding R$^{3+}$ ion and O$^{2-}$ in 8-fold coordination \cite{Shannon:a12967}. 

\subsection{Non-magnetic Electronic structure}

 We first investigate the evolution of the non-magnetic (NM) electronic structure of LaNiO$_2$ for independent in-plane and out-of-plane lattice constant reductions and then correlate these with the change in R across the lanthanide series. Fig. \ref{fig1} shows the non-magnetic band structure for LaNiO$_2$ at the experimental lattice parameters compared with that at 4, and 8\% reduction of in-plane (right) and out-of-plane (left) lattice constants (a zoomed-in version of the band structures around the Fermi level is shown in Fig. \ref{figa2}). In all cases, a single Ni-d$_{x^2-y^2}$ band crosses the Fermi level. In addition, there are La-$5d$ bands that give rise to two electron pockets that self-dope the Ni-d$_{x^2-y^2}$ band, as determined in previous work on LaNiO$_2$ and NdNiO$_2$ \cite{pickett, prx, arita, Been2020TheoryOR, doped_MI, ES_112}. The pockets at $\Gamma$ 
and A have predominant La-$d_{z^2}$ and La-$d_{xy}$ character, respectively.

\begin{figure}
\includegraphics[width=0.9\columnwidth]{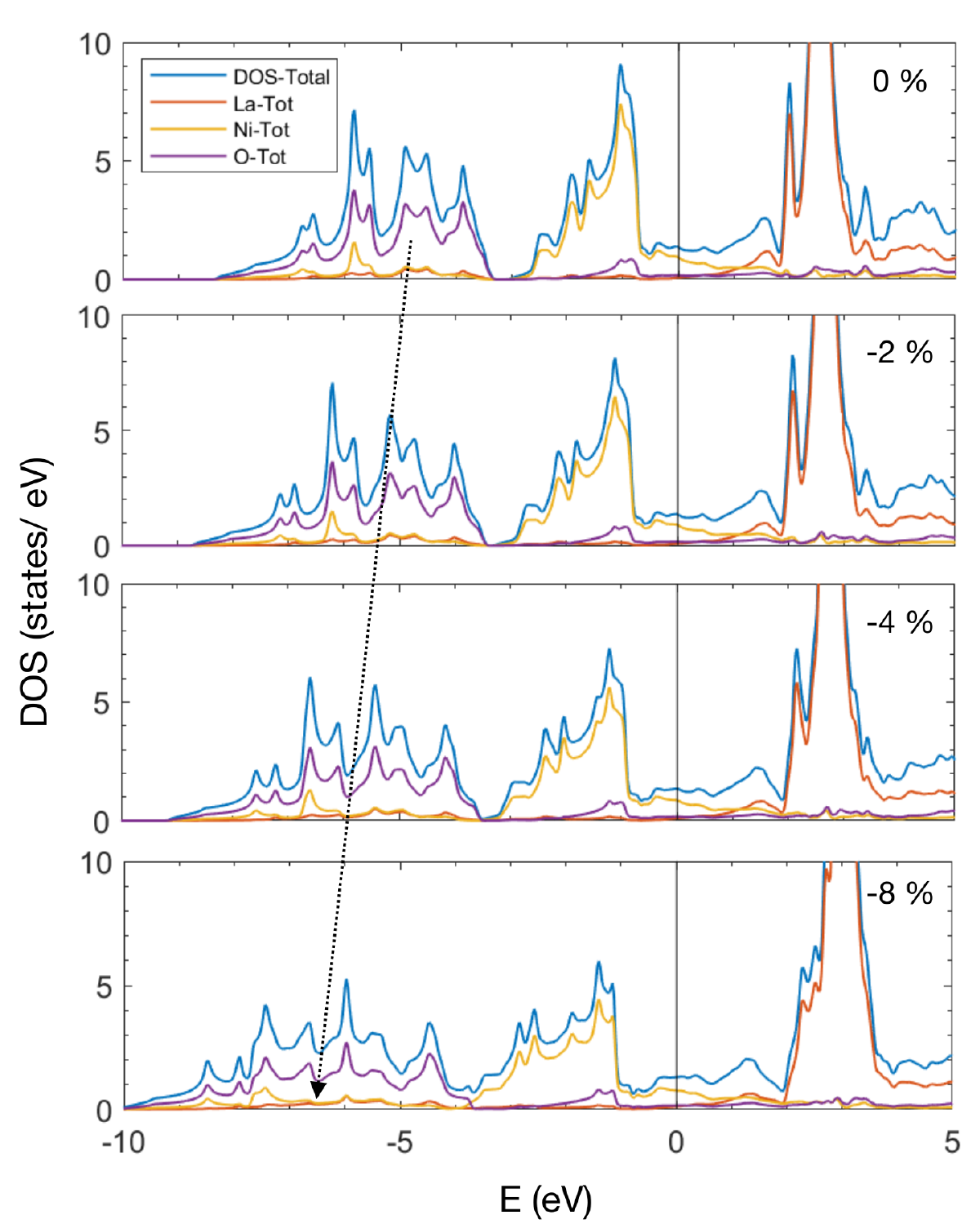}
\caption{Total, La-d+f, Ni-d and O-p atom-resolved density of states for GGA non-magnetic calculations in LaNiO$_2$ for decreasing in-plane lattice parameter with respect to the experimentally reported $a$ value. For each plot, the out-of-plane lattice parameter, $c$, is held constant.}
\label{fig2}
\end{figure}

\begin{figure}
\includegraphics[width=\columnwidth]{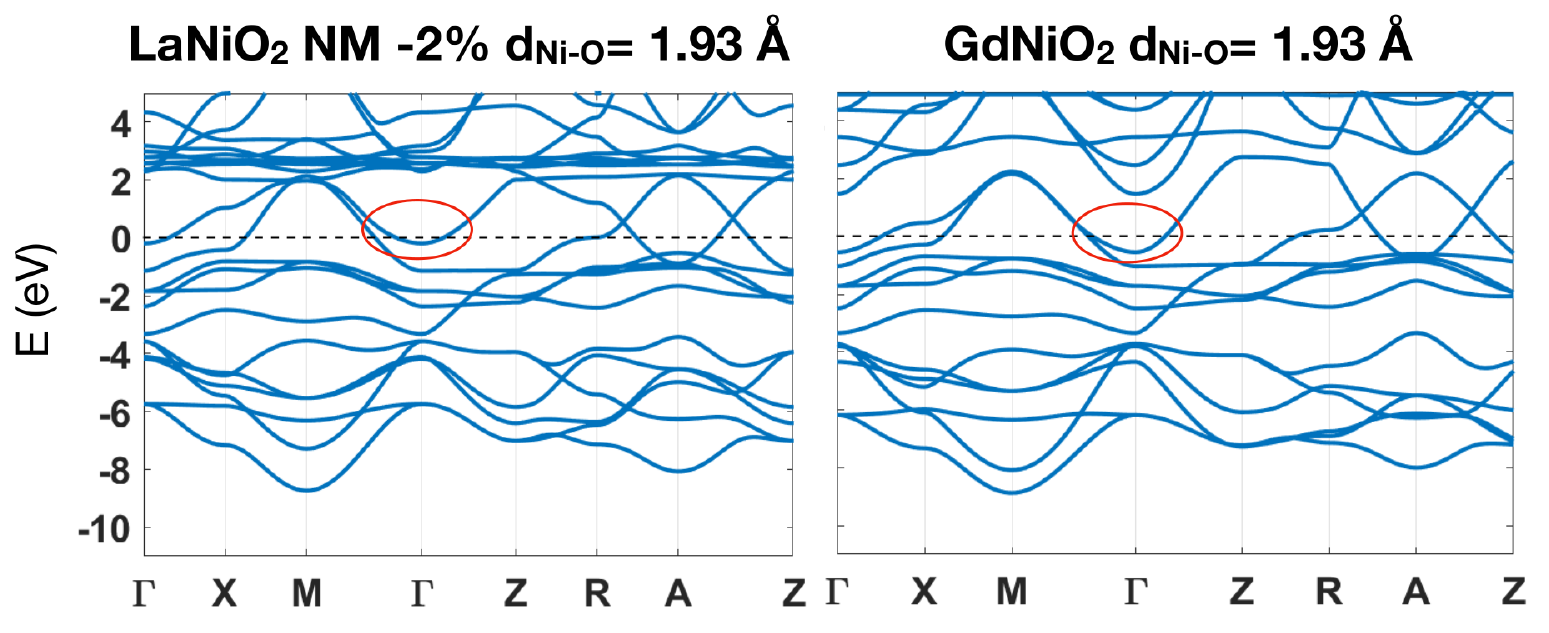}
\caption{GGA non-magnetic bandstructure for 1) LaNiO$_2$ with a 2.4\% reduction of the in-plane lattice constant with respect to the experimental value (left), and 2) GdNiO$_2$ (right). Both systems are compared as they have the same Ni-O in-plane distance. The red oval encloses the R-d electron pocket at $\Gamma$. }
\label{fig3}
\end{figure}

Even though upon reducing the lattice parameters  there are still bands of Ni-d$_{x^2-y^2}$ and La-d$_{xy}$+d$_{z^2}$ character crossing the Fermi level (E$_F$), important differences arise upon changing the in-plane lattice parameter. In contrast, altering the out-of-plane lattice constant does not give rise to appreciable differences in the bandstructure (see Fig. \ref{fig1}). Specifically, upon reducing the in-plane lattice constant (or Ni-O in-plane distance) the most important modifications in the bandstructure are: 1) The increase in size of the La-d pockets both at A and at $\Gamma$.  An increase in the size of the R-d pocket at A has been linked to the increase of the hopping between the
interstitial and R-d$_{xy}$ orbitals in Ref. \onlinecite{arita_2} when studying trends on different hypothetical ${d}^{9}$ layered nickelates.
 2) The increase in the bonding-antibonding splitting of the Ni-d$_{x^2-y^2}$ and O-p$\sigma$ states, noticeable at the M point  (2 eV increase upon an 8\% in-plane lattice parameter reduction with respect to the experimental value). 3) The increase in bandwidth of the d$_{x^2-y^2}$ band crossing the Fermi level (from 3 eV for the experimental lattice constants to 4.6 eV for an 8\% reduction of $a$). These trends are the expected ones: as the in-plane Ni-O distance decreases, the effective d$_{x^2-y^2}$ nearest neighbor hopping $t$ increases, and so does the bandwidth. The La-d pockets increase to compensate for the overall upward shift of the Ni-d$_{x^2-y^2}$ band as the lattice parameters are reduced. 

An important difference between 112 nickelates and cuprates, that has been highlighted before,\cite{pickett, prx} is the larger charge transfer energy  ($\Delta=E_{d}-E_{p}$) in the former. The $\Delta$ value derived from on-site energies of the Wannier functions for LaNiO$_2$ is $\sim$ 4.4 eV \cite{prx} whereas typical cuprate values are $\sim$ 2 eV \cite{weber}. This would put 112 nickelates within the Mott-Hubbard regime in the Zaanen-Sawatzky-Allen clasification \cite{zsa} in contrast to cuprates, a prototypical example of charge transfer insulator due to the high degree of p-d hybridization \cite{berciu}. Fig. \ref{fig2} shows the orbital resolved density of states for different in-plane lattice parameter reductions within LaNiO$_2$ with respect to the experimental value. The most important change in Fig. \ref{fig2} upon a Ni-O in-plane distance reduction (adding up to effects 1-3 mentioned above) is 4) is the shift of the O-p centroid to lower energies ($\sim$ 2 eV shift for an 8\% $a$ reduction, 1 eV for a 4\% $a$ reduction) while the Ni-d centroid does not significantly move. This last point is very important since it gives rise to a decrease in the degree of p-d hybridization (or an increase in charge transfer energy) as the Ni-O in-plane distance is reduced. This is consistent with Ref. \onlinecite{Been2020TheoryOR} that reports a reduction in p-d hybridization as R changes from La to Lu (i.e. as the lattice parameter is reduced) as inferred from the degree of oxygen admixture in the lower Hubbard band.

Effects 1-4 in the electronic structure derived from a simple in-plane lattice parameter reduction in LaNiO$_2$, can be correlated directly to R changes across the lanthanide series by comparing electronic structures that correspond to the same Ni-O in-plane distance. For example, a $\sim$ 2.4\%  in-plane lattice constant reduction in LaNiO$_2$ can be compared directly with the bandstructure of GdNiO$_2$ as they correspond to the same Ni-O distance of 1.93 \AA~. Fig. \ref{fig3} shows the band structures for these two systems. The bandwidth of the d$_{x^2-y^2}$ bands is identical in the two cases (extending from $\sim$ 2 to -1 eV), as is the energy range of the O-p (from $\sim$ -4 to -9 eV) and Ni-d (from $\sim$ 2 to -3.7 eV) states. Also identical is the position of the O-p centroid. The explicit change in R only gives rise to a slightly different  size of the R-d pocket at $\Gamma$ (marked in red). The same reasoning can be extended to the other R ions (see Appendix Figures \ref{figa3}, \ref{figa4}, and \ref{figa5} for more details).

In essence, all the changes in the electronic structure of RNiO$_2$ nickelates linked to a change in R (namely the bandwidth, size of the R-d pockets and charge transfer energy), can be mimicked by a simple change of in-plane lattice constant corresponding to the same Ni-O distance. In contrast, a large change in out-of-plane lattice constant has negligible effects in the electronic structure. Hence, we conclude that the relevant control parameter for the  electronic structure of 112 nickelates is the Ni-O in-plane distance. A change in R has, per se, negligible effects in the electronic structure, an expected outcome given their chemical similarity. It is simply the Ni-O in-plane distance reduction a change in R gives rise to (as the size of R gets reduced from La to Lu) that has profound effects in the electronic structure. This result contrasts with Ref. \onlinecite{Been2020TheoryOR} in that we find no significant effects that we can ascribe to a specific rare-earth or to the change in out-of-plane lattice constants.

\begin{figure}
\includegraphics[width=\columnwidth]{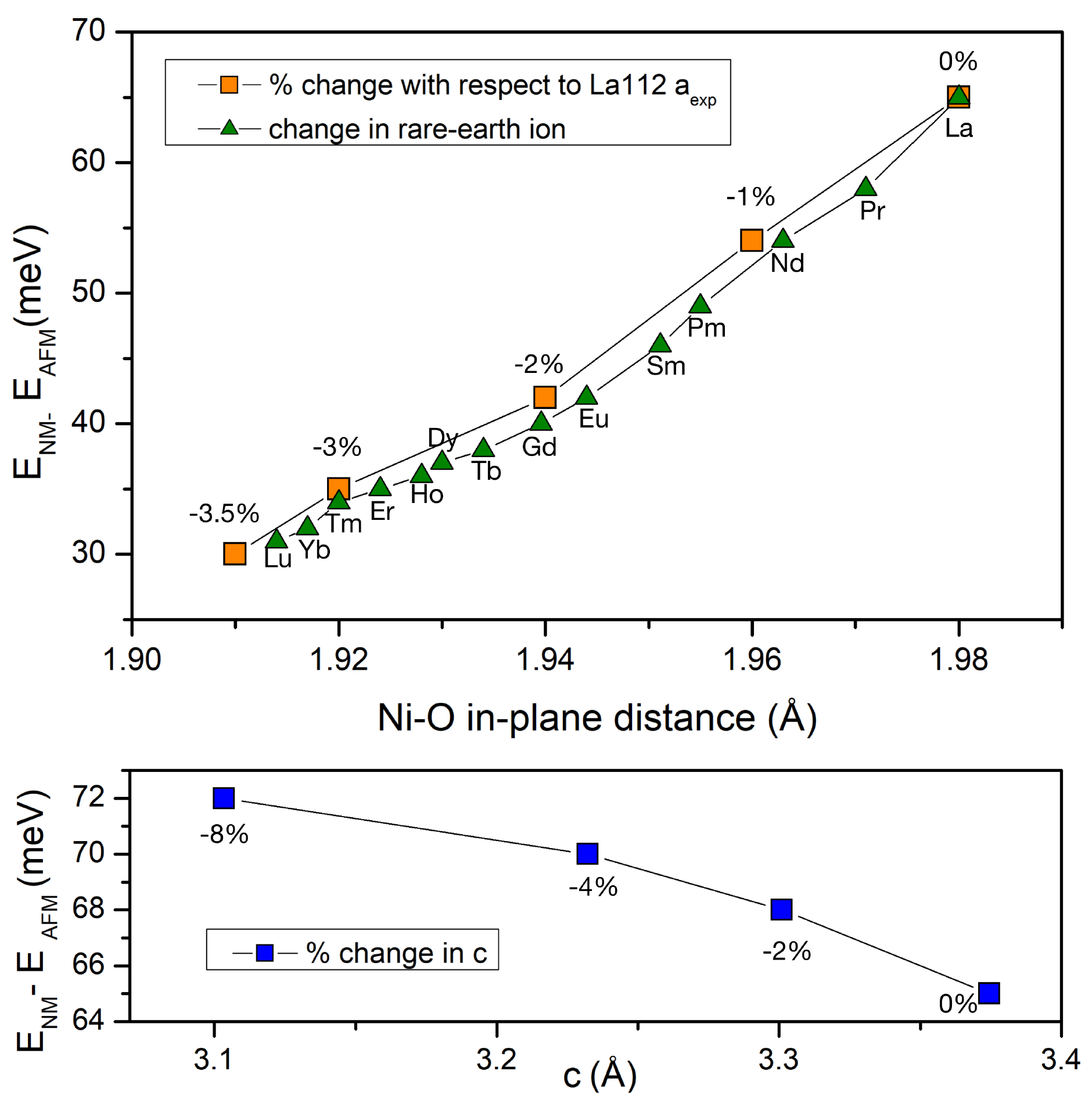}
\caption{Energy difference within GGA between a C-type AFM state and a NM state for 112 nickelates as a function of: 1) Ni-O in-plane distance corresponding to changes in R and to a lattice parameter reduction with respect to the experimentally reported $a$ value for LaNiO$_2$ (top panel) and 2) out-of-plane lattice parameter reduction with respect to the experimentally reported $c$ value for LaNiO$_2$  (bottom panel).}
\label{fig4}
\end{figure}

\begin{figure}
\includegraphics[width=\columnwidth]{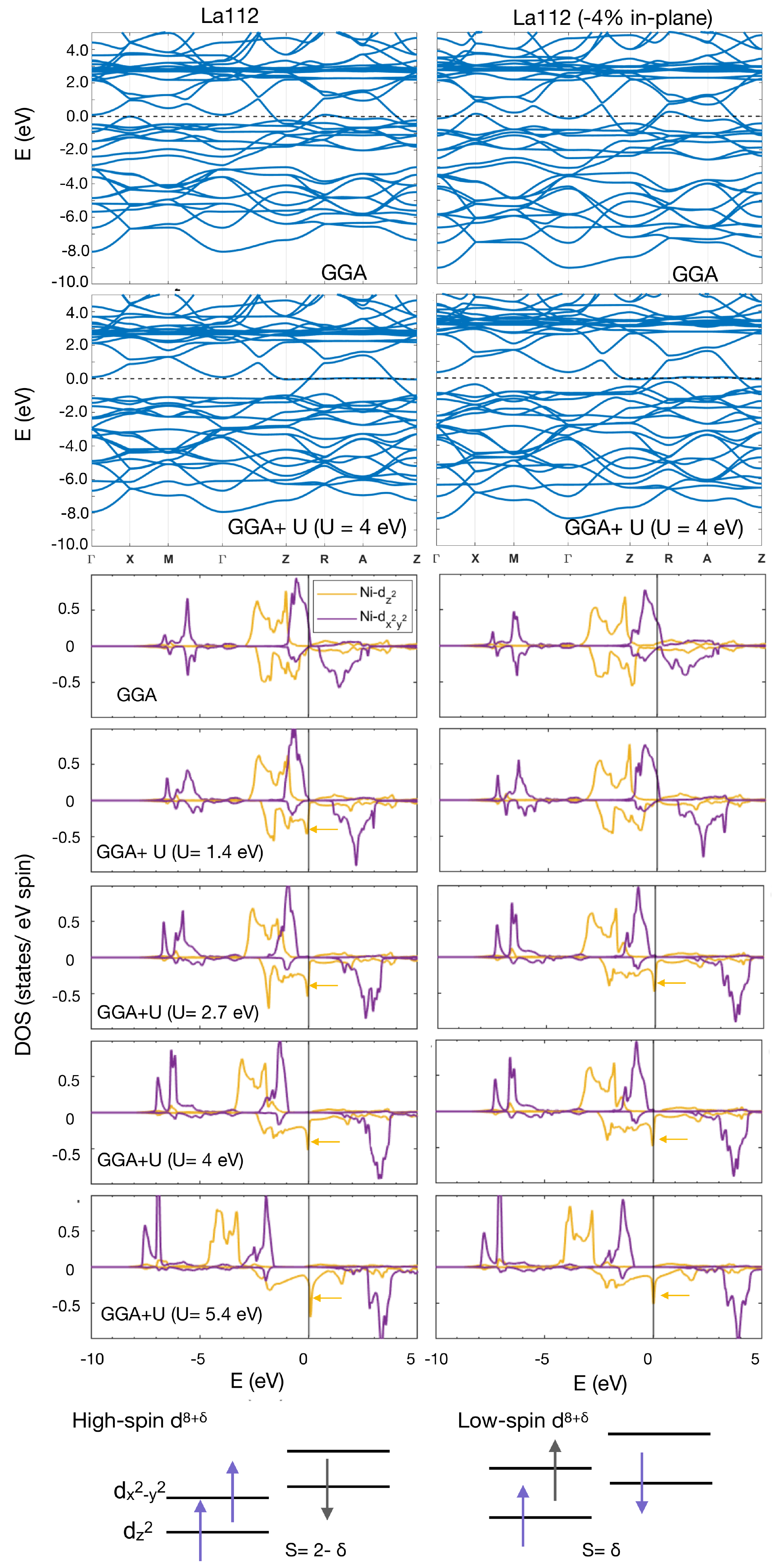}
\caption{Top panels: C-type AFM band structure of LaNiO$_2$ for the experimental lattice parameters (left) and for a 4\% $a$ reduction (right)  within GGA, and GGA+U (4 eV). Middle panels: Evolution of the Ni-d$_{x^2-y^2}$ and Ni-d$_{z^2}$ DOS for LaNiO$_2$ in the AFM state within GGA and GGA+U upon increasing U (U= 1.4-5.4 eV). Note the 1D van Hove singularity pinned at the Fermi level, with pure d$_{z^2}$ character, for U $\geq$ 1.4 eV marked by the yellow arrows. Bottom panels: Schematic representation of the energy level diagrams for high-spin Ni$^{2+\delta}$(d$^{8+\delta}$) ion (left) and low-spin (right) in a square-planar environment. The t$_{2g}$-like levels are fully occupied so we do not show them here. The gray arrow reflects the partial occupation of that particular orbital.} 
\label{fig5}
\end{figure}

\subsection{Spin-polarized calculations}

Strong antiferromagnetic correlations are considered a key ingredient in cuprates. In contrast, there is no experimental evidence for antiferromagnetic order in 112 nickelates to date \cite{ikeda, ikeda2, hayward}. 
 Recent DMFT calculations in 112 nickelates point to the importance of a multiorbital picture with low-energy relevance of a
flat Ni-d$_{z^2}$ band, and the existence of high spin Ni d$^8$ \cite{lechermann, lechermann2, Petocchi2020NormalSO, Werner2020NickelateSM}. DFT calculations using C-type AFM order in NdNiO$_2$ are consistent with this picture in that they also show a flat Ni-d$_{z^2}$ band arising at the Fermi level   \cite{pickett3}. This work concludes that this flat band  may be related to instabilities
that could limit AFM order at low
temperature and preclude the AFM
phase with the formation of an AFM spin-liquid state that forms the
platform for superconductivity \cite{pickett3}.  In order to determine the importance and robustness of this flat-band feature in the AFM state for other R ions we analyze, using spin-polarized calculations, different magnetic configurations for RNiO$_2$ (R= La-Lu) and correlate these trends with in-plane and out-of-plane lattice parameter  changes  in LaNiO$_2$ (as done in the previous section for NM calculations).

 Both a C-type and a G-type AFM state are more stable than a NM state for LaNiO$_2$ from first principles. A G-type AFM state is the ground state within GGA (with a small energy difference of 5 meV/Ni with respect to the C-type AFM state, and 70 meV/Ni with respect to the NM state). However, the C-type AFM state becomes more stable than a G-type one within GGA+U for U $\geq$ 1.4 eV. This is consistent with the results in Ref. \onlinecite{liu} (see Fig. \ref{figa1} for more details). Attempts to stabilize a FM state give rise to a reduced magnetic moment of $\sim$ 0.2 $\mu_B$ at the GGA level, less stable than any of the AFM states by $\sim$ 0.72 meV/Ni. Note that all these energy differences are small. 

 As the C-type AFM state is the ground state for a reasonable U value  for these  metallic nickelates (U $\sim$ 4-5 eV) \cite{Sakakibara_arxiv2019, arita}, this is the AFM order we employ to show magnetic trends in all other RNiO$_2$ materials to draw a systematic comparison.  We note that Ref. \onlinecite{Been2020TheoryOR} describes the evolution of the electronic structure with U in the G-type AFM state instead. G-type AFM order is also the focus of Ref. \onlinecite{npj_112} for NdNiO$_2$.
 
 The GGA-energy difference  between the C-type AFM state and the NM one for RNiO$_2$ (R= La-Lu) is shown in Fig. \ref{fig4} as a function of R and in-plane lattice parameter change. For every R ion and for every lattice parameter used, the ground state of the 112 materials keeps on being AFM. However, as the in-plane lattice parameter/size of R decreases, the energy difference between the AFM and NM state decreases as well (see Fig. \ref{fig4}).  Based on these considerations, in contrast to Ref. \onlinecite{Been2020TheoryOR}, we anticipate that the tendency towards AFM should be suppressed in R112 as the size of R/Ni-O in-plane distance is reduced, consistent with the concomitant increase in bandwidth described above. 
 Importantly, the energy difference linked to the in-plane lattice parameter change can be matched almost exactly to that related to a change in R, in agreement with the main conclusion drawn from the NM calculations in the previous section: the change in Ni-O in-plane distance controls the electronic structure of these systems and the R change is not relevant other than for the Ni-O distance change it carries. To further reinforce that the Ni-O in-plane distance is the control parameter in 112 materials, the energy difference between a NM and an AFM state is also shown for an out-of-plane lattice constant reduction. A large 8 \% reduction of the out-of-plane lattice constant gives rise to a very small energy difference across the series ($\sim$ 7 meV) and does not produce significant changes in the electronic structure.

We now turn to the nature of the AFM state. A simple ionic count for 112 nickelates gives a Ni valence of Ni$^{1+}$: d$^9$, as mentioned above. However, the self-doping effect effectively gives rise to a d$^{8+\delta}$ ion (with a large $\delta$). In a square
planar environment (with a large crystal field splitting between the d$_{x^2-y^2}$ and d$_{z^2}$ bands) two possible spin states can occur (see Fig. \ref{fig5}). If the crystal field splitting within the e$_g$ states is larger than Hund's rule coupling, a low spin (LS) state develops (with S= $\delta$/2 and a moment ($\delta$)$\mu_B$ per nickel). If the Hund's rule coupling is larger, a high spin (HS) state would be more
stable (with S= (2-$\delta$)/2 and a moment (2-$\delta$)$\mu_B$ per nickel). The HS and LS states lead to different properties not only in terms of the moments but also in terms of the electronic structure. In particular, for a HS state, Ni-d$_{z^2}$ states are relevant in the vicinity of the Fermi level, in contrast to the more cuprate-like scenario of a LS state, with only a Ni-d$_{x^2-y^2}$ being relevant. Hence, the careful analysis of HS vs LS states in R112 nickelates is very important.  In this context, it has recently been reported that layered oxychalcogenides A$_2$NiO$_2$Ag$_2$Se$_2$(A = Sr,Ba), with a NiO$_2$ square lattice and Ni d$^8$, may exhibit HS S=1 Ni$^{2+}$\cite{layered_nickelate_exp}, or a Ni on-site 'off-diagonal singlet' in which both e$_g$ orbitals are singly occupied but with
Kondo-like oppositely spin-directed singlets   \cite{lee_layered}.

Figure \ref{fig5} shows the AFM-GGA band structure for LaNiO$_2$ at the experimental lattice parameters and for a 4\% in-plane lattice parameter reduction. Both band structures look similar: the derived state is metallic with a Ni-d$_{x^2-y^2}$ band crossing the Fermi level and a La-d pocket around Z. The derived Ni-magnetic moments are consistent with LS-Ni d$^{8+\delta}$ ($\sim$ 0.7 $\mu_B$ for the experimental lattice constants, $\sim$ 0.6 $\mu_B$ for a 4\% in-plane lattice parameter reduction). The only differences upon reducing the in-plane lattice parameter are: 1) a slight reduction of the Ni-magnetic moment and 2) an increase in bandwidth, as expected. The same trends are observed for AFM calculations upon a change in R across the lanthanide series (see Figs. \ref{figa6}-\ref{figa8}). 

Within GGA+U, the magnetic moments and electronic structure of LaNiO$_2$ at small U value ($\leq$ 1 eV) are very similar to those obtained within GGA, with a Ni-d$_{x^2-y^2}$ and La-d band crossing the Fermi level. At a U value of 1.4 eV, a flat d$_{z^2}$ band emerges at E$_F$ as shown in Fig. \ref{fig5} that depicts the Ni-d$_{z^2}$ and Ni-d$_{x^2-y^2}$ DOS upon increasing U and shows the corresponding bandstructure at U= 4 eV. In this bandstructure the flat band can be clearly observed along the Z-R-A-Z direction (for the bandstructures at other U values see Appendix Fig. \ref{figa9}). We find that this feature is robust in DFT for LaNiO$_2$ (and other Rs, see below), in agreement with what Ref. \onlinecite{pickett3} reports for NdNiO$_2$ within DFT, and also in agreement with recent DMFT work  \cite{lechermann, lechermann2, Petocchi2020NormalSO, Werner2020NickelateSM}. We note that the La-d states appear well below the Fermi level (see  Appendix Fig. \ref{figa10}). As U increases, so does the value of the Ni magnetic moment from the 0.7 $\mu_B$ obtained within GGA for the experimental lattice constants, up to $\sim$ 1.3 $\mu_B$ within GGA+U for the highest $U$ value shown of 5.4 eV. This change in the electronic structure is consistent with a LS-to-HS transition with increasing U (even though this behavior had been observed before in DFT calculations \cite{pickett3,prx}, it had never been ascribed to a LS-to-HS Ni-d$^{8+\delta}$ transition).

The metallic character of the AFM state obtained for La112, the LS-to-HS transition upon increasing U, and the flat d$_{z^2}$-band feature are robust upon reduction of the in-plane lattice parameter (as shown in Fig. \ref{fig5}) and, more importantly, upon a change in R (see Figs. \ref{figa6}-\ref{figa9}).  All in all, this peculiar AFM state with a flat d$_{z^2}$ band pinned at the Fermi level is stable from first-principles calculations for every R and every lattice parameter. Hence, we anticipate the instabilities  limiting  AFM  order at  low  temperature  described in Ref. \onlinecite{pickett3} might be in action in all other parent RNiO$_2$ materials if they can be synthesized.

\section{conclusions}

Using DFT calculations, we have analyzed the evolution of the electronic structure and magnetic properties of RNiO$_2$ nickelates as R changes across the lanthanide series. By  correlating these changes with in-plane and out-of-plane lattice parameter reductions in LaNiO$_2$ (the extreme member of the series), we determine that the electronic and magnetic responses of infinite-layer nickelates are governed by the in-plane Ni-O distance. In contrast, changes in out-of-plane lattice constant have negligible effects. The non-cuprate like properties reported for La/NdNiO$_2$ persist for other rare-earth ions, i. e. involvement of R-bands at the Fermi level and large charge transfer energy. The ground state for RNiO$_2$ materials from first-principles calculations for every R and lattice parameter studied is AFM and metallic, even though the tendency towards AFM is suppressed as the size of R gets reduced from La to Lu. In contrast to cuprates, this AFM state is characterized by  multi-orbital character with a flat Ni-d$_{z^2}$ band pinned at the Fermi level (in addition to the naively expected d$_{x^2-y^2}$ states) that enables the formation of high-spin Ni states. It was recently suggested \cite{pickett3} that for NdNiO$_2$  this flat band can make the system unstable with respect to charge, spin, and lattice orders, limiting AFM order at low temperature. The robustness of this flat d$_{z^2}$ band for every R (and lattice parameter) in RNiO$_2$ nickelates suggests that the same instabilities may arise across the lanthanide series if other members can be synthesized.

\section{Acknowledgments}
 We acknowledge fruitful discussions with V. Pardo. ASB acknowledges NSF-DMR grant 1904716. We acknowledge the ASU Research Computing Center for HPC resources.

\appendix

\section{GGA and GGA+U Electronic structure RNiO$_2$}

Figure \ref{figa1} shows the  energy difference   between C-type and G-type AFM ordering ($E_{C}-E_{G}$) for LaNiO$_2$ within GGA and GGA+U at different values of U. Within GGA, $E_{C}-E_{G}>0$ so G-type ordering is more stable. However, for U $\geq$ 2, the energy difference is negative, thus the C-type AFM order is more stable. In addition, both C- and G-type AFM states are more stable than a NM state for GGA and for GGA+U at all U-values.

 Figure \ref{figa2} depicts the  non-magnetic  band structures for LaNiO$_2$  shown in Fig. \ref{fig1} in a zoomed-in  energy range so that the behavior around the Fermi level can be ascertained more easily. The changes to the band structure are readily apparent for a reduction in the in-plane lattice parameter (increase in Ni-d$_{x^2-y^2}$ bandwidth and size of La-d pockets), whereas a reduction in the out-of-plane lattice parameter produces negligible changes.

Fig. \ref{figa3} shows the non-magnetic GGA band structure plots of RNiO$_2$ for twelve different R ions with associated Ni-O distances. Comparatively, Figure \ref{figa4} shows the non-magnetic GGA band structure plots of LaNiO$_2$ for different in-plane lattice parameters, ranging from the experimental one, to  an 8\% reduction, along with associated Ni-O distances. One can see the similarities between band structures of comparable Ni-O distance and also the same trend of increasing bandwidth as the Ni-O distance is decreased for different R ions (as in Fig. \ref{figa3}) and for decreasing in-plane lattice parameter (as in Fig. \ref{figa4}). 

Figure \ref{figa5} depicts the non-magnetic R-d+f, Ni-d and O-p atom-resolved density of states plots for eight different R ions in R112. One can notice the increase in charge-transfer energy as the size of R decreases with increasing atomic number across the lanthanide series. The same trends are observed in LaNiO$_2$ for decreasing in-plane lattice parameter, as shown in Fig. \ref{fig3} in the main text. 

Figures \ref{figa6}, \ref{figa7}, and \ref{figa8} depict the AFM band structure plots for RNiO$_2$ for twelve R ions withing GGA and GGA+U (U=4 eV, J=0.68 eV). The bandwidth can be seen to increase for decreasing lattice parameter, and a flat Ni-d$_{z^2}$ band appears pinned at the Fermi energy along Z-R-A-Z within GGA+U. For each value of U, the Ni magnetic moment decreases as the size of R decreases with increasing atomic number across the lanthanide series.

Figure \ref{figa9} shows the AFM band structure plots of LaNiO$_2$ for different in-plane lattice parameters (experimental (0\%) , 2\%, and 4\% reduction) within GGA and GGA+U (U= 2.7, 4, and 5.4 eV, J=0.68 eV). The same flat Ni-d$_{z^2}$ band appears pinned at the Fermi energy along Z-R-A-Z for non-zero values of U. For each value of U, the Ni magnetic moment decreases with decreasing in-plane lattice parameter.

Figure \ref{figa10} shows the La-d orbital-resolved density of states for LaNiO$_2$ for a C-type AFM order at different in-plane lattice parameters (0\% and -4\% reduction) within GGA and GGA+U (U= 1.4, 2.7, 4, and 5.4 eV).

\renewcommand{\thefigure}{A\arabic{figure}}
\setcounter{figure}{0}

\begin{figure*}
\includegraphics[width=\columnwidth]{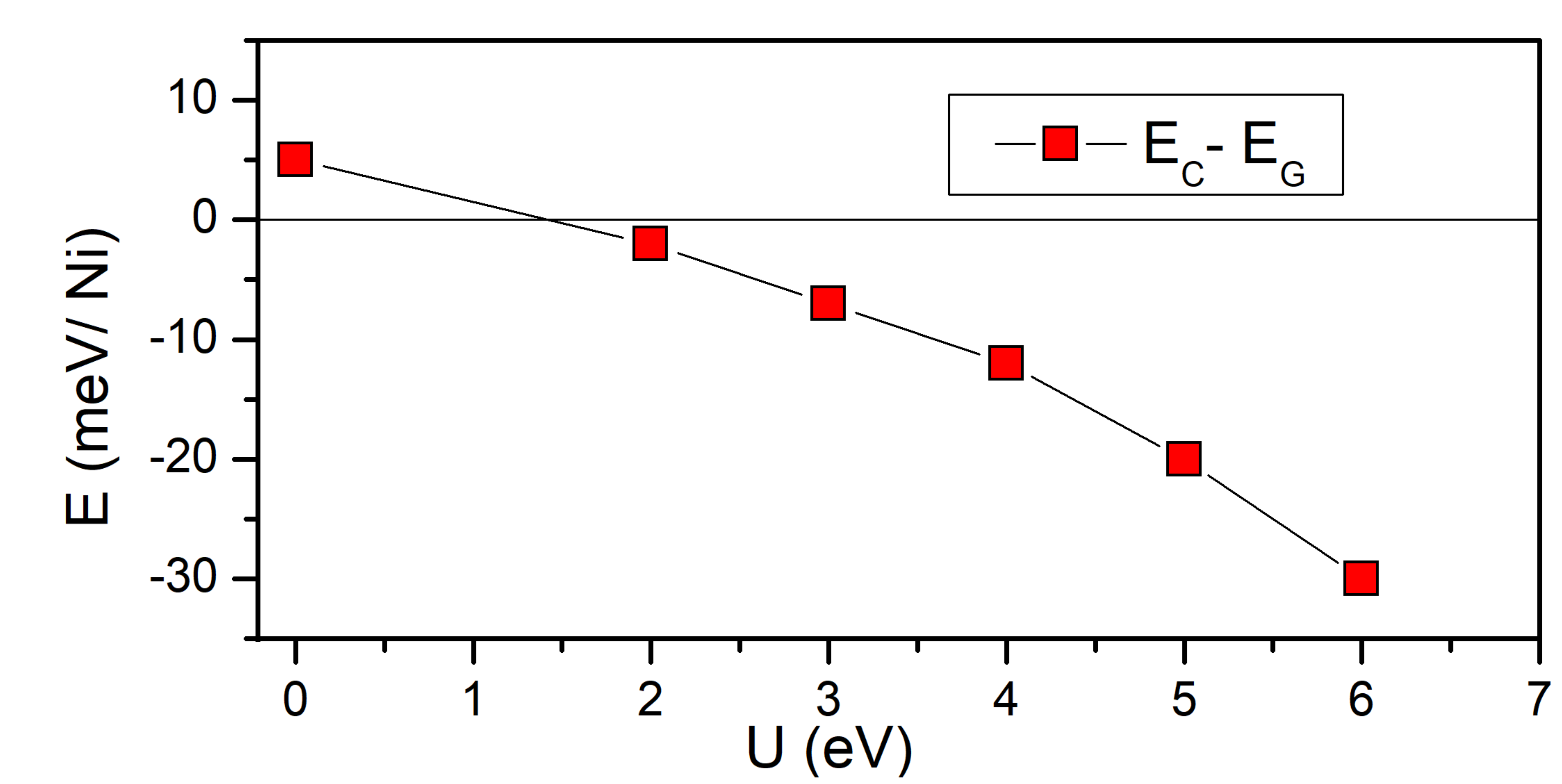}
\caption{Energy difference between the two lowest-energy magnetic states (G-type AFM, and C-type AFM) for LaNiO$_2$ within GGA and GGA+U as a function of U (U= 2-6 eV). Negative indicates C-type is stable, positive indicates G-type is stable.}
\label{figa1}
\end{figure*}

\begin{figure*}
\includegraphics[width=2\columnwidth]{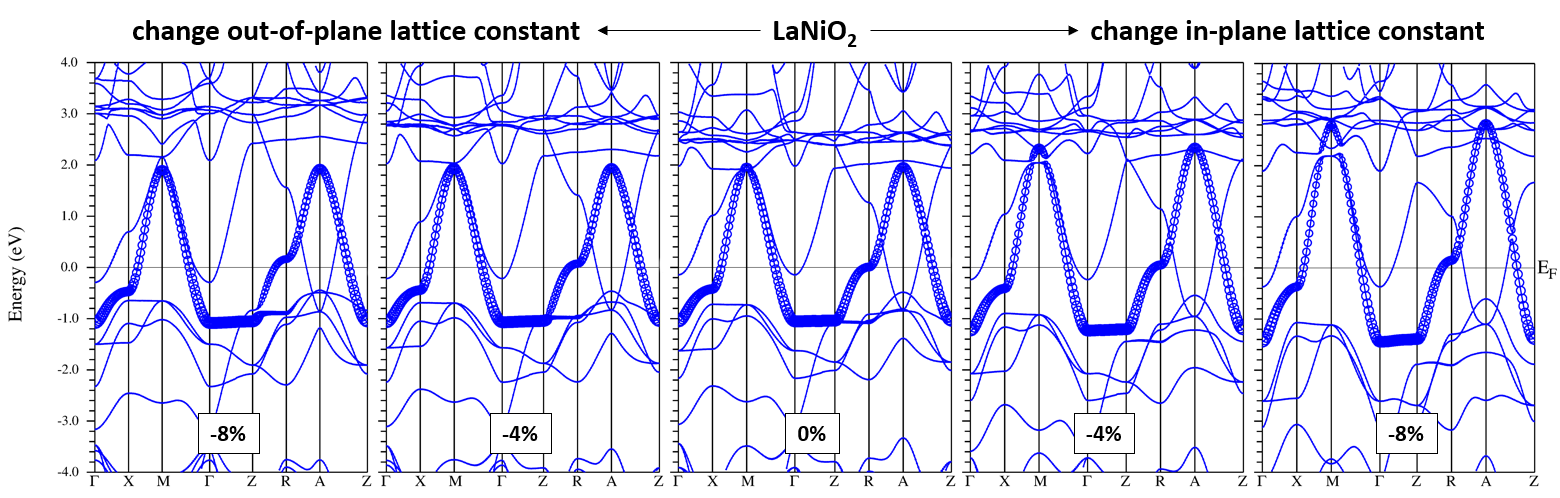}
\caption{GGA non-magnetic LaNiO$_2$ band structure with band character plot (Ni-d$_{x^2-y^2}$ highlighted) at the experimental lattice constants (0\%, central panel), upon reduction of in-plane (right panels) and out-of-plane (left panels) lattice parameters. This is a reproduction of the band structures in Fig. \ref{fig1} zoomed-in around the Fermi energy.}
\label{figa2}
\end{figure*}

\begin{figure*}
\includegraphics[width=2\columnwidth]{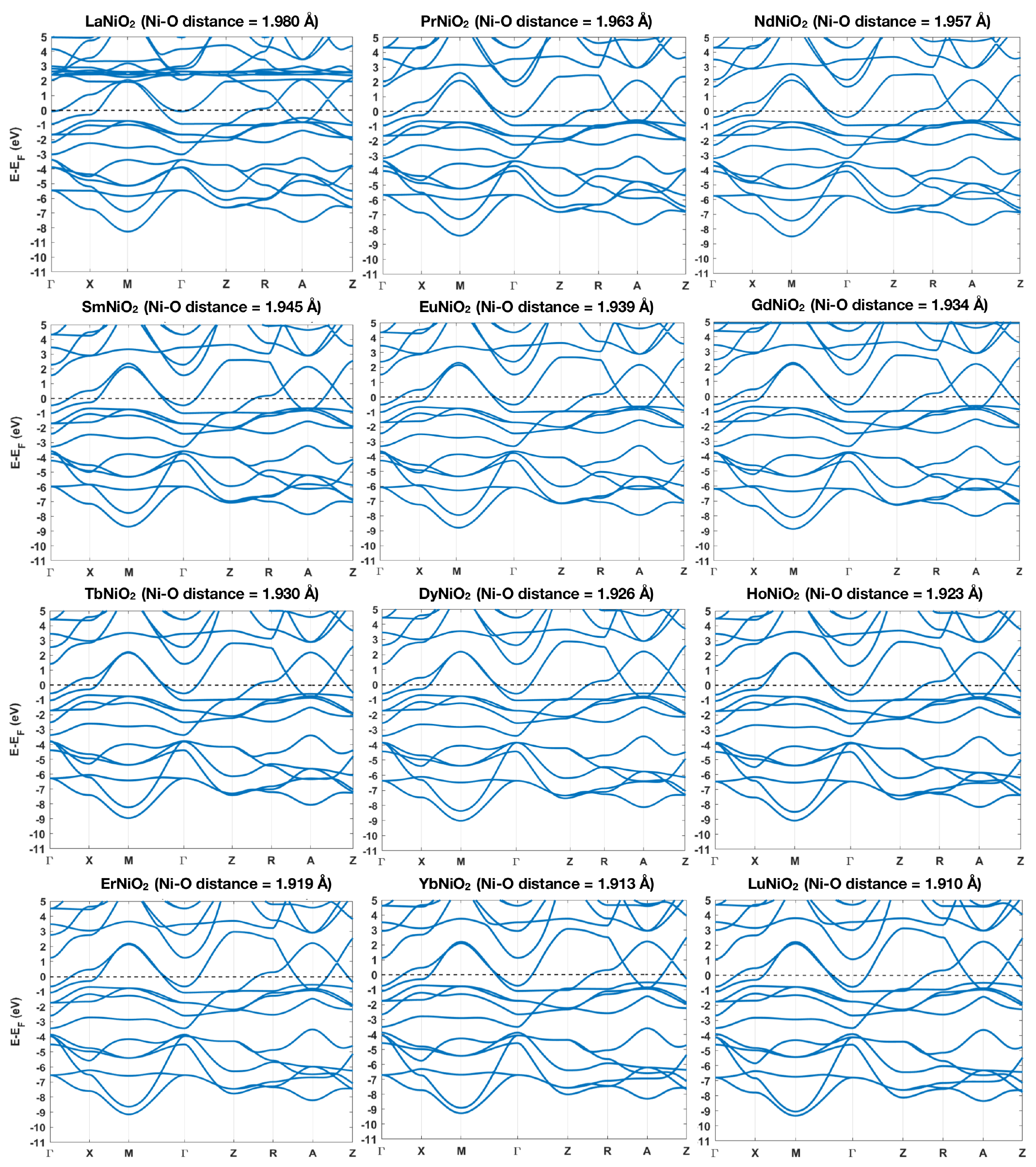}
\caption{Non-magnetic GGA band structure plots for  RNiO$_2$ (R= La-Lu).}
\label{figa3}
\end{figure*}

\begin{figure*}
\includegraphics[width=1.8\columnwidth]{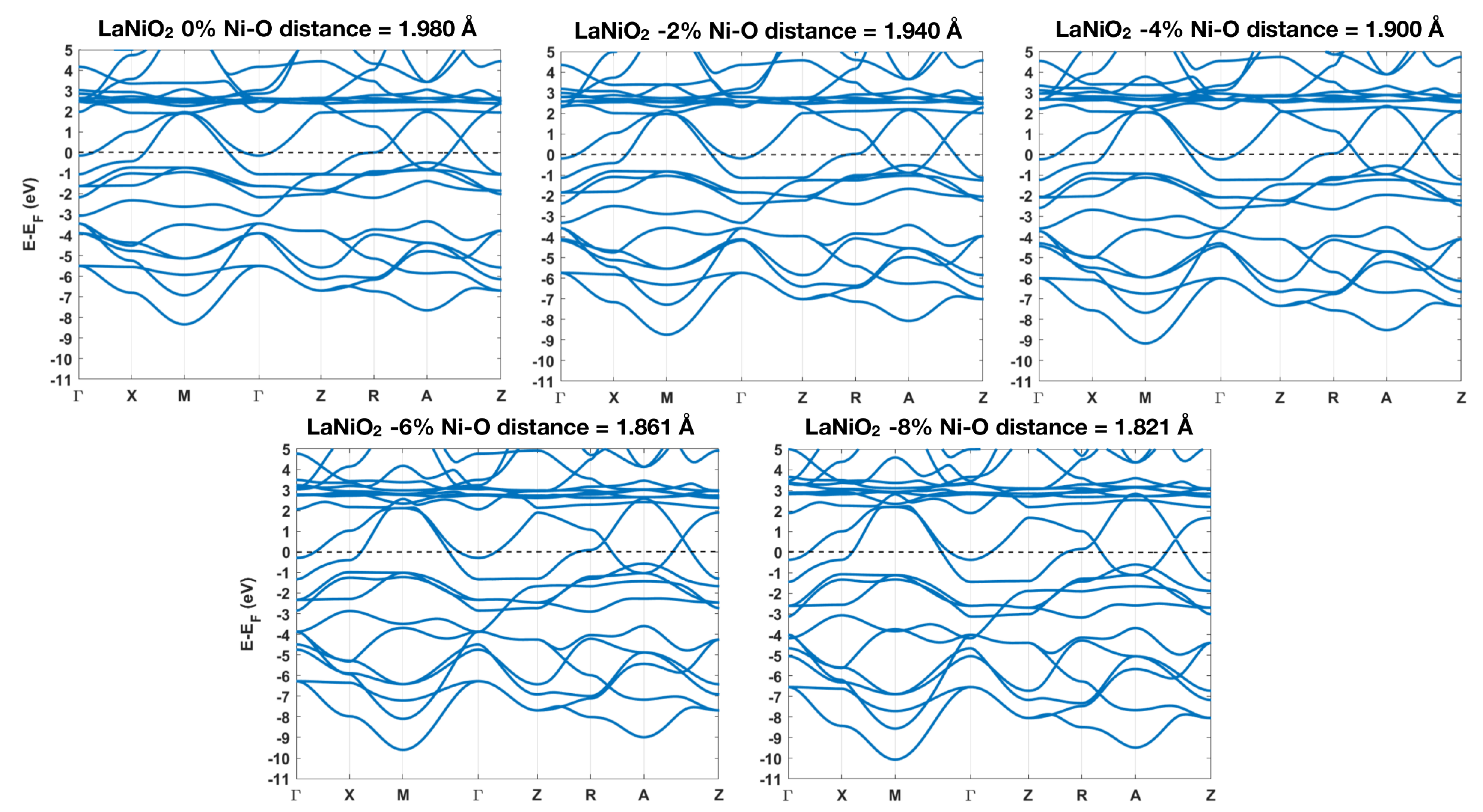}
\caption{Non-magnetic GGA band structure plots for LaNiO$_2$ at different in-plane lattice parameter reductions with respect to the experimental one.}
\label{figa4}
\end{figure*}

\begin{figure*}
\includegraphics[width=1.6\columnwidth]{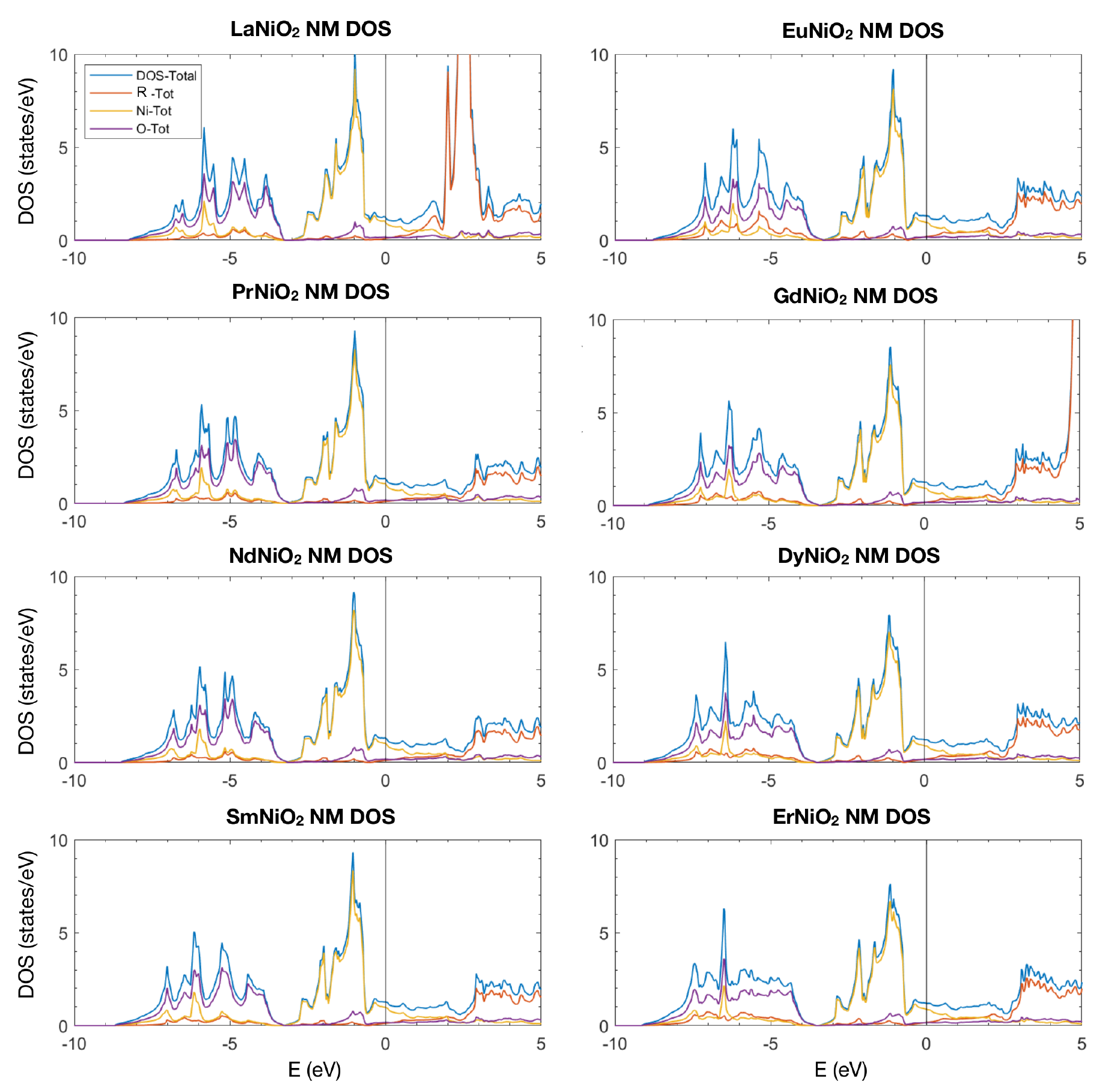}
\caption{Total, R-d+f, Ni-d and O-p atom-resolved density of states for different R112.}
\label{figa5}
\end{figure*}

\begin{figure*}
\includegraphics[width=1.6\columnwidth]{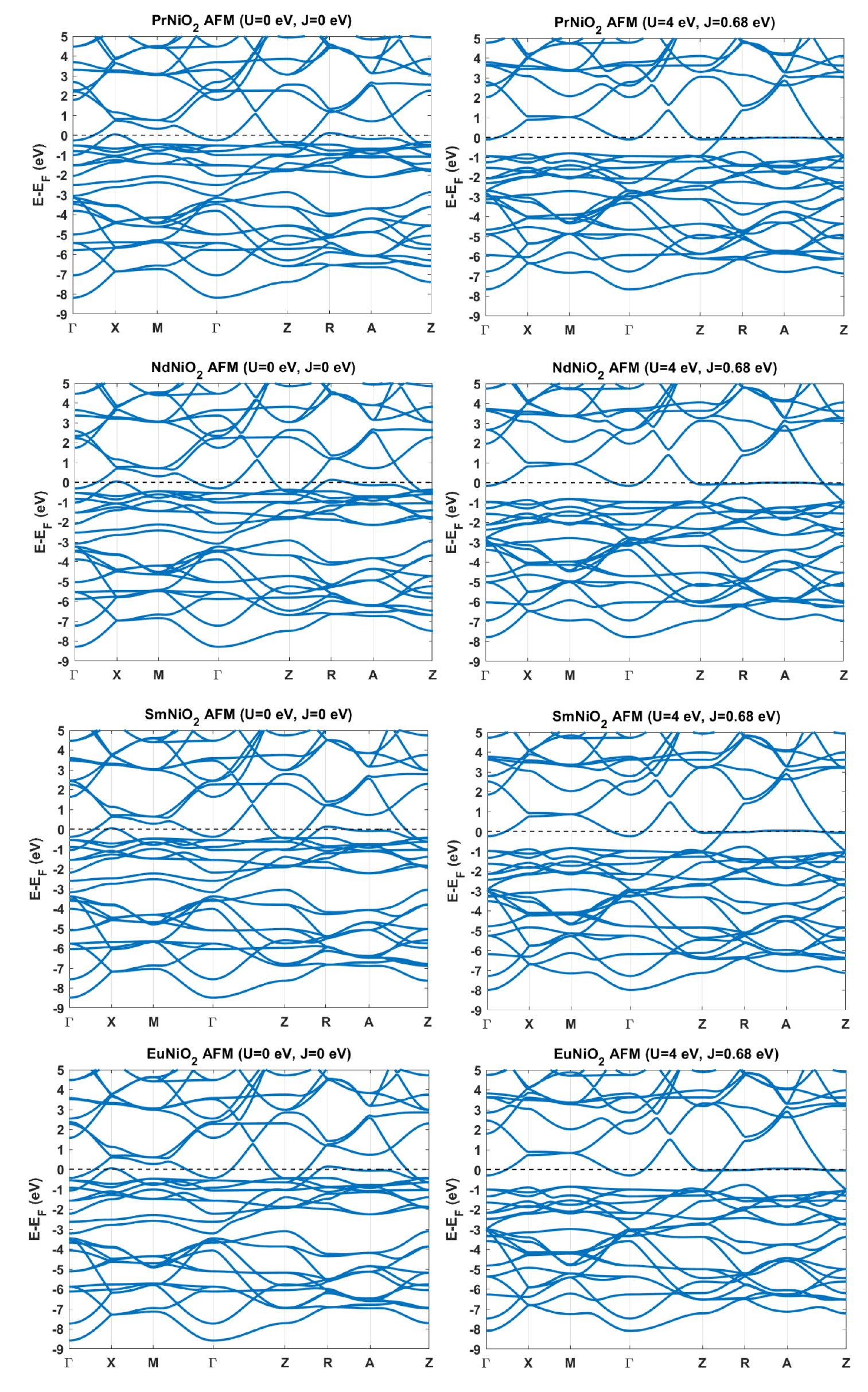}
\caption{AFM GGA (left) and GGA+U (U= 4 eV, right) bandstructure for RNiO$_2$. Notice the flat band of Ni-d$_{z^2}$ character pinned at E$_F$ along Z-R-A-Z.  
}
\label{figa6}
\end{figure*}

\begin{figure*}
\includegraphics[width=1.6\columnwidth]{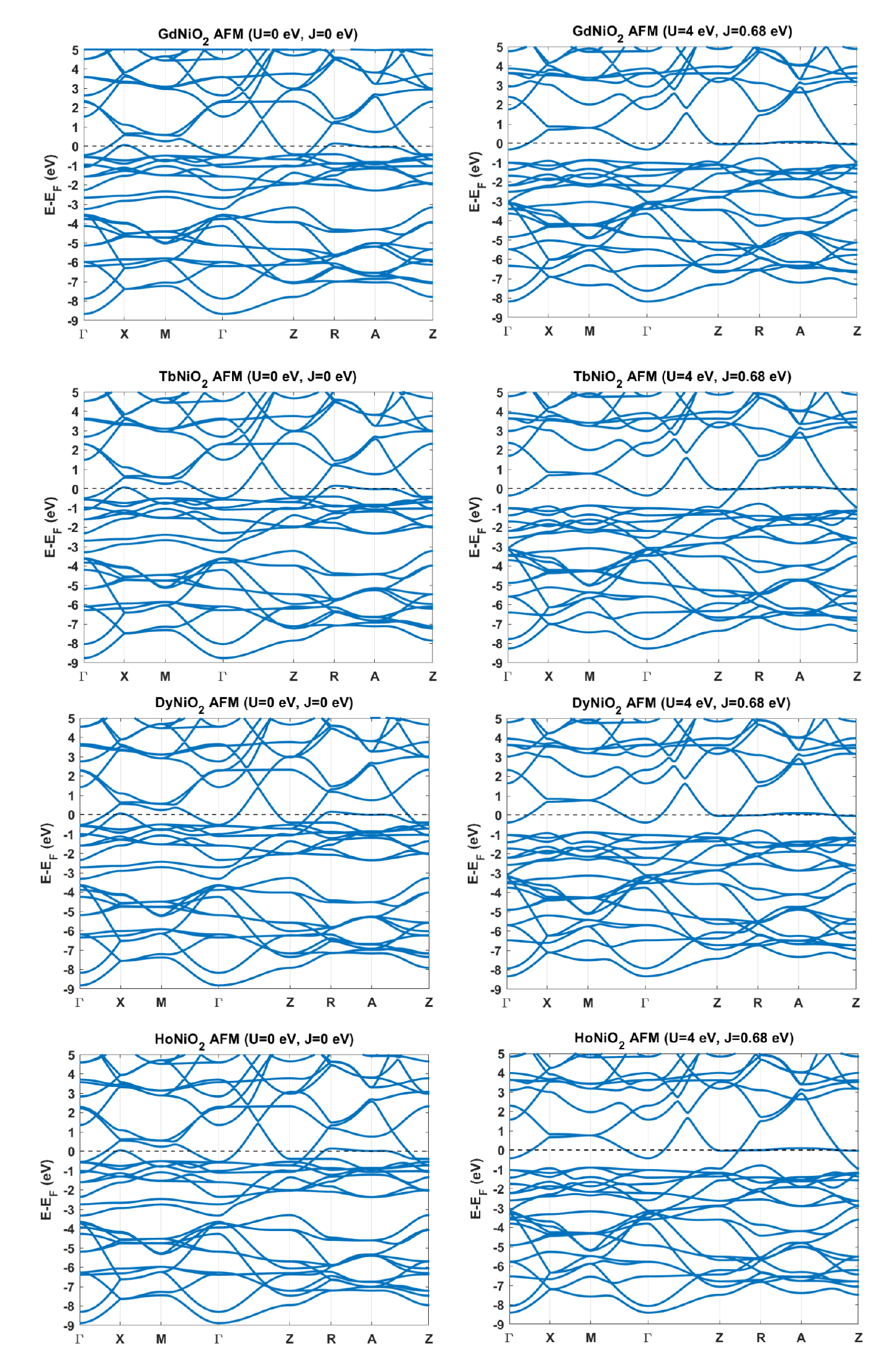}
\caption{(continued) AFM GGA (left) and GGA+U (U= 4 eV, right) bandstructure for RNiO$_2$. Notice the flat band of Ni-d$_{z^2}$ character pinned at E$_F$ along Z-R-A-Z.
}
\label{figa7}
\end{figure*}

\begin{figure*}
\includegraphics[width=1.7\columnwidth]{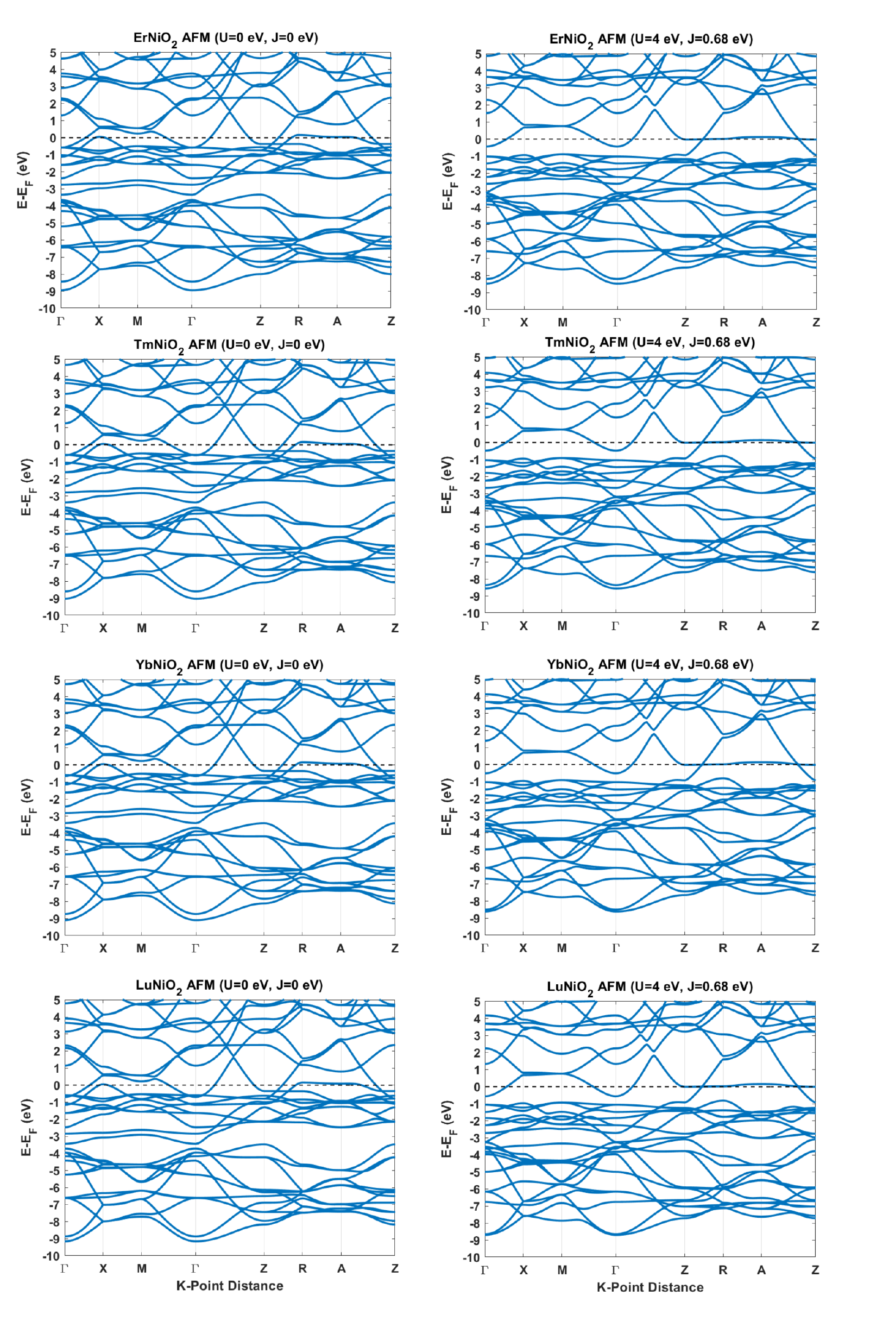}
\caption{(continued) AFM GGA (left) and GGA+U (U= 4 eV, right) bandstructure for RNiO$_2$. Notice the flat band of Ni-d$_{z^2}$ character pinned at E$_F$ along Z-R-A-Z. 
}
\label{figa8}
\end{figure*}

\begin{figure*}
\includegraphics[width=1.8\columnwidth]{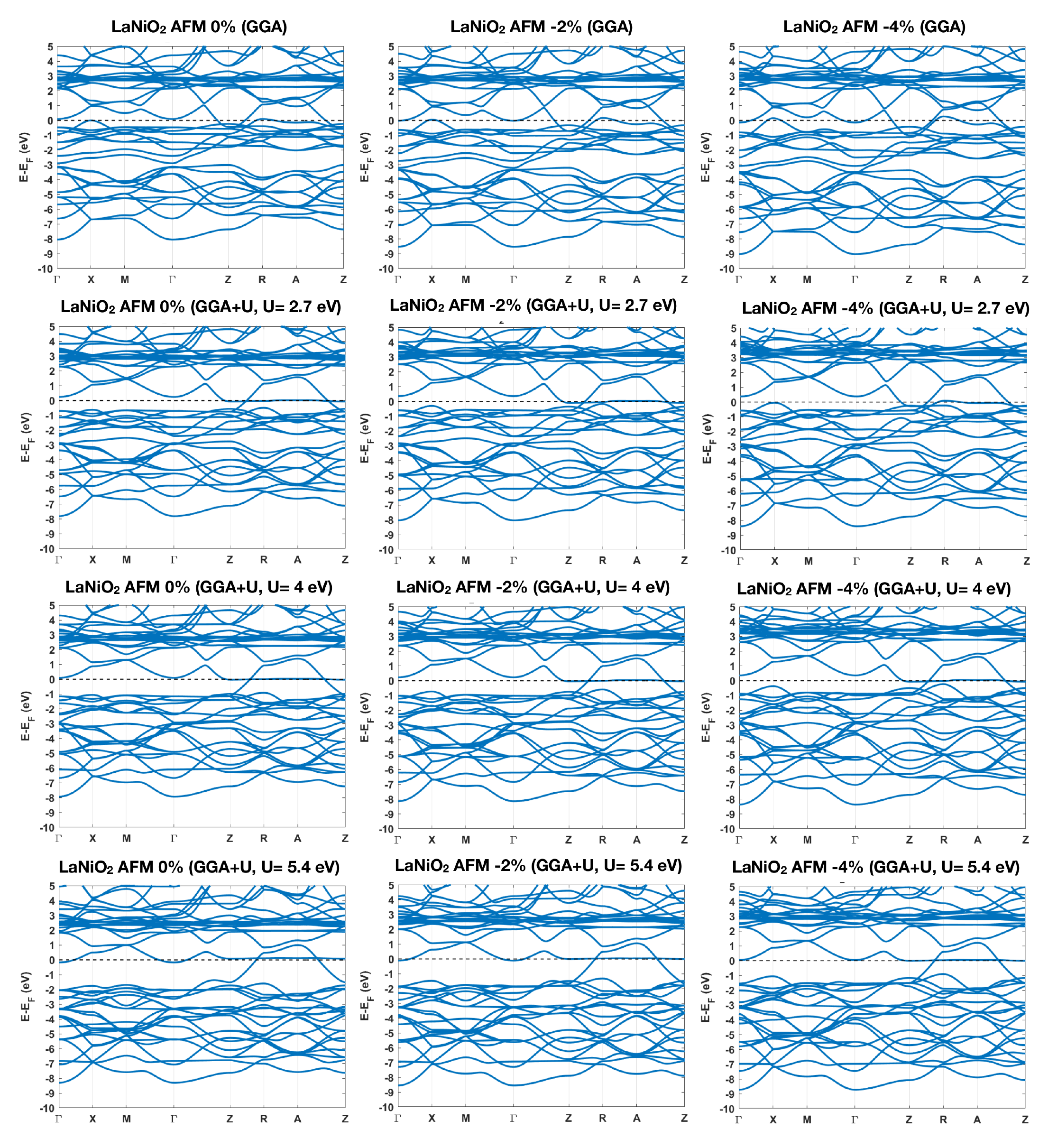}
\caption{AFM GGA and GGA+U (U= 2.7, 4 and 5.4 eV) band structure plots LaNiO$_2$ at different in-plane lattice parameters. Notice the flat band of Ni-d$_{z^2}$ character pinned at E$_F$ along Z-R-A-Z. 
}
\label{figa9}
\end{figure*}

\begin{figure*}
\includegraphics[width=2\columnwidth]{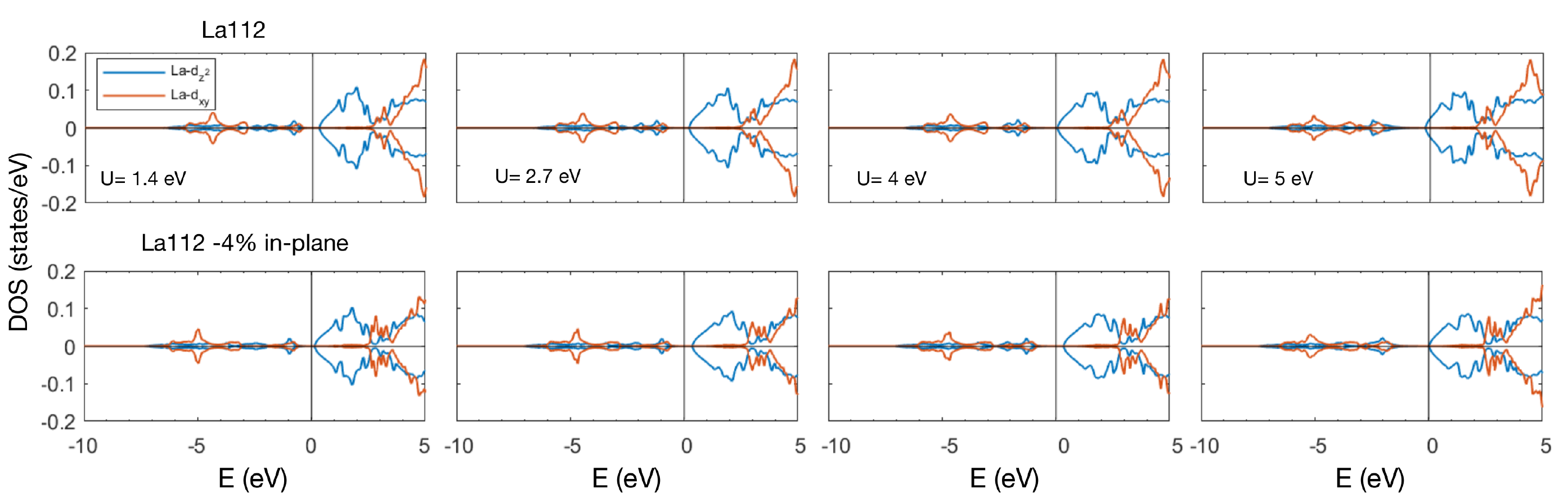}
\caption{AFM GGA+U (U= 1.4 to 5.4 eV) La-d orbital resolved DOS for LaNiO$_2$ at different in-plane lattice parameters. }
\label{figa10}
\end{figure*}

\end{document}